\documentclass[a4paper]{jpconf}
\usepackage{graphicx, color, subfigure}
\usepackage{bm}
\usepackage{soul}
\setstcolor{blue}

\newcommand{\wo}{\omega_{osc}}
\newcommand{\wa}{\omega_a}
\newcommand{\wk}{\omega_{k}}
\newcommand{\wj}{\omega_{j}}

\newcommand{\wc}{\omega_{cut}}
\newcommand{\bk}{{\bf k}}
\newcommand{\bE}{{\bf E}}

\begin{document}
\title{Effect of boundaries on vacuum field fluctuations and radiation-mediated interactions between atoms}

\author{F Armata$^1$, S Butera$^2$, G Fiscelli$^3$, R Incardone$^{4,5}$, V Notararigo$^6$, \\
R Palacino$^3$, R Passante$^{3,7}$, L Rizzuto$^{3,7}$, S Spagnolo$^3$}

\address{$^1$ QOLS, Blackett Laboratory, Imperial College London, London SW7 2AZ, UK}
\address{$^2$ SUPA, Institute of Photonics and Quantum Sciences, Heriot-Watt University, Edinburgh, UK}
\address{$^3$ Dipartimento di Fisica e Chimica, Universit\`{a} degli Studi di Palermo, Via Archirafi 36, I-90123 Palermo, Italy}
\address{$^4$ 4th Institute for Theoretical Physics, Universit\"{a}t Stuttgart, 70569 Stuttgart, Germany}
\address{$^5$ Max Planck Institute for Intelligent Systems, 70569 Stuttgart, Germany}
\address{$^6$ Department of Physics and Astronomy, University College London, Gower Street, London WC1E 6BT, UK}
\address{$^7$ INFN, Laboratori Nazionali del Sud, I-95123 Catania, Italy}

\ead{lucia.rizzuto@unipa.it}

\begin{abstract}
In this paper we discuss and review several aspects of the effect of boundary conditions and structured environments on dispersion and resonance interactions involving atoms or molecules, as well as on vacuum field fluctuations. We first consider the case of a perfect mirror, which is free to move around an equilibrium position and whose mechanical degrees of freedom are treated quantum mechanically. We investigate how the quantum fluctuations of the mirror's position affect vacuum field fluctuations for both a one-dimensional scalar and electromagnetic field, showing that the effect is particularly significant in the proximity of the moving mirror. This result can be also relevant for possible gravitational effects, since the field energy density couples to gravity. We stress that this interaction-induced modification of the vacuum field fluctuations can be probed through the Casimir-Polder interaction with a polarizable body, thus allowing to detect the effect of the mirror's quantum position fluctuations. We then consider the effect of an environment such as an isotropic photonic crystal or a metallic waveguide, on the resonance interaction between two entangled identical atoms, one excited and the other in the ground state.
We discuss the strong dependence of the resonance interaction with the relative position of the atomic transition frequency with the gap of the photonic crystal in the former case, and with the cut-off frequency of waveguide in the latter.

\end{abstract}

\section{Introduction}

A peculiar aspect of the quantum theory is the presence of unavoidable quantum fluctuations of the fields, even in the ground state of the system \cite{Milonni94}. Field fluctuations have striking consequences in quantum electrodynamics; for example:  the Lamb shift, the anomalous magnetic moment of the electron, Casimir and Casimir-Polder forces, vacuum polarization \cite{CompagnoPassante95,Buhmann12}. In quantum field theory, the vacuum state is not \emph{void}, but contains virtual particles or quanta which give a sort of structure to the empty space \cite{Aitchison85}  and have observable effects on matter. On the other hand, the presence of matter, i.e. microscopic objects such as atoms or molecules as well as macroscopic objects, can change the physical features of vacuum fluctuations and ultimately of the vacuum state. Thus, macroscopic objects as fixed or moving metallic or dielectric plates or structured materials such as photonic crystals or photonic crystals waveguides can modify vacuum field fluctuations and the properties of the vacuum \cite{SimpsonLeonhardt15,Lamoreaux05}, yielding striking observable effects such as the Casimir effect, that is a force between two neutral conducting plates in the vacuum, and Casimir-Polder forces \cite{Casimir48,CasimirPolder48,Milonni07}. Moreover, this allows us to modify and tailor radiative processes that are related to vacuum fluctuations, for example spontaneous emission \cite{Purcell46,Kleppner81,Yablonovitch87}, radiative corrections \cite{WangKivshar04} and radiation-mediated interaction between atoms and molecules (for example, van der Waals/Casimir-Polder \cite{HaakhScheel15,HoodGoban16,MessinaPassante08,ShahmoonMazetz15} and resonance interactions \cite{ElGanainyJohn13,IncardoneFukuta14}).

In this paper we shall discuss and review some aspects related to the role of fixed or moving boundaries on vacuum field energy densities and Casimir interactions, as well as the influence of a photonic bandgap environment or a metallic waveguide on the resonance (radiation-mediated) interaction between two entangled identical atoms, one excited and the other in the ground state.
Specifically, we will first consider vacuum field energy densities near a boundary that is free to move and whose mechanical degrees of freedom are treated quantum mechanically, thus subjected to quantum fluctuations of its position. Compared to the case of a fixed configuration, the effect of the boundary's position fluctuations
is mainly concentrated in the very proximity of the boundary itself \cite{ButeraPassante13,ArmataPassante15} and yields a smearing out of
the otherwise divergent field energy density at the position of the boundary \cite{FordSvaiter98,BartoloButera15}. This fact not only affects the Casimir-Polder force on a polarizable body placed near the wall, but, in principle, can be relevant from a gravitational point of view, because the energy density couples to the gravitational field.
We will then consider a photonic bandgap environment, in particular a photonic crystal, and discuss how it can alter the resonance interaction between two atoms embedded in the crystal \cite{IncardoneFukuta14,NotararigoPassante16}. Finally, we briefly mention the case of the resonance interaction between two atoms placed inside a metallic cylindrical waveguide.

This paper is organized as follows. In section \ref{Fluctwall} we consider a one-dimensional cavity with a mobile wall described quantum mechanically and discuss how quantum fluctuations of its position affect the field energy density inside the cavity and the Casimir-Polder interaction energy with a polarizable body placed in the proximity of the fluctuating wall, as well as the observability of this effect. In section \ref{PBG} we discuss the effect of an external environment such as a photonic crystal or a metallic cylindric waveguide on the resonance interaction between two entangled atoms, mediated by the radiation field. Last Section is devoted to our conclusive remarks.

\section{Vacuum fluctuations near a fluctuating wall and Casimir-Polder interactions}
\label{Fluctwall}

We first consider a one-dimensional massless scalar field confined between two perfectly reflecting mirrors placed at $x=0$ and $x=L(t)$. The mirror at $x=0$ has a fixed position, while we allow the other mirror of mass $M$ to move in space, assuming it bounded by a harmonic potential of frequency $\wo$ around an equilibrium position $L_0$. We assume Dirichlet boundary conditions for the scalar field operator at the boundaries, i.e. $\phi (0,t)=\phi(L(t),t)=0$. The mechanical degrees of freedom of the moving mirror are treated quantum mechanically; this permits us to include quantum position fluctuations of the mirror. We find that their effect is particularly relevant for small mirror's masses, such as those attainable in modern optomechanics experiments \cite{AspelmeyerKippenberg14}. Since the mobile mirror can oscillate, an interaction between the cavity field modes and the mirror's mechanical degrees of freedom (phonons) is generated, as well as an effective interaction between field modes, mediated by the moving mirror. Our system is described by the following effective Hamiltonian, valid for a small displacement of the moving mirror from its equilibrium position \cite{Law94,Law95}
\begin{equation}
 H = \hbar \wo b^\dag b +\hbar \sum_k \omega_k a_k^\dag a_k
 - \sum_{kj}C_{kj} \Big\{ \left( b + b^\dag \right)
{\cal N} \! \left[ \left( a_k + a_k^\dag \right) \left(  a_j + a_j^\dag \right) \right] \Big\}  ,
\label{Hamiltonian1D}
\end{equation}
where $a_k, \, a_k^\dagger$ are annihilation and creation operators of the cavity field modes relative to the equilibrium position $L_0$ with wavenumber $k$, $b, \, b^\dagger$ are annihilation and creation operators relative to the mobile wall (mechanical excitations), ${\cal N}$ is the normal ordering operator, and
\begin{equation}
C_{kj}=(-1)^{k+j}\left(\frac \hbar 2\right)^{3/2}\frac 1{L_0 \sqrt{M}}\sqrt{\frac{\omega_k \omega_j}\wo}
\label{Coefficient1D}
\end{equation}
is the effective coupling constant between the moving mirror and the field.  In our 1D model, the field modes are equally spaced in frequency and the allowed frequencies are
$\omega_j = ck_j$, where  $k_j=j\pi /L_0$, with $j$ an integer number.

We now consider the field energy density inside the cavity. The corresponding operator is

\begin{equation}
\mathcal{H}(x)=\frac 12 \left[\frac 1{c^2} {\dot{\phi}}^2(x)+\left( \frac {d {\phi (x)}}{dx} \right)^2 \right] .
\label{Hamiltonian}
\end{equation}
After subtraction of the (divergent) energy density that would be present even in absence of the wall, in the case of fixed walls its average value on the ground state
$| \{ 0_k\}\rangle$ of the field, is $\langle\{0_k\}|\mathcal{H}|\{0_k\}\rangle = - \pi  c \hbar /(24 L_0^2)$,
yielding a constant value inside the cavity. At the position of the two fixed walls, extra divergent terms are present, as in the case of the vacuum electromagnetic field fluctuations \cite{FordSvaiter98,SopovaFord02,BartoloPassante12}. In the case of fluctuating boundaries, such divergences can be smeared out by an average on the probability distribution function of the mirror's position  \cite{FordSvaiter98,BartoloButera15}.
We now consider how the field energy density changes when we allow one of the mirrors to move around its equilibrium position, using the Hamiltonian (\ref{Hamiltonian1D}). Due to the interaction term in (\ref{Hamiltonian1D}), the bare ground state $|\{0_k\},0\rangle$ of the field-mobile wall system is not an eigenstate of the total Hamiltonian. The true (dressed) ground state can be obtained by first-order perturbation theory, and it contains an admixture of states with a mechanical excitation of the mirror and pairs of virtual excitations of the cavity field. At the lowest order in the mirror-field interaction, the dressed ground state is
\begin{equation}
| g \rangle=| \{ 0_k \} ,0 \rangle+\sum_{kj} D_{kj} | \{1_k,1_j \} ,1 \rangle ,
\label{dressedstata1D}
\end{equation}
where
\begin{equation}
D_{kj}=(-1)^{k+j}\frac 1{L_0} \sqrt{\frac {\hbar \wk \wj}{8 M \wo}} \left( \frac 1{\wo+\wk+\wj} \right) .
\label{coeffstate1D}
\end{equation}
In Eq. (\ref{dressedstata1D}), the elements inside the curly bracket in the state vectors refer to field excitations, while the other element refers to phonons, that is excitations of the wall's mechanical degrees of freedom.

We can now evaluate the average value of the \emph{renormalized} energy density on the dressed state (\ref{dressedstata1D}), namely after subtraction of its space-independent value in the absence of the walls given above, obtaining \cite{ButeraPassante13}
\begin{equation}
\langle g | \mathcal{H}(x) | g \rangle_R = \frac{\hbar^2}{2L_0^3 M \wo} \sum_{j\ell r} (-1)^{\ell +r} \cos [(q_\ell -q_r)x]
\frac {\omega_\ell \omega_j\omega_r}{(\wo+\omega_\ell +\omega_j)(\wo +\omega_r+\omega_j)}  ,
\label{averenergydensity1D}
\end{equation}
where $q_\ell = \ell \pi /L_0$, with $\ell$ an integer number. To evaluate (\ref{averenergydensity1D}), it is necessary to introduce an upper cutoff frequency $\wc$, related to a sort of plasma frequency of the cavity walls. To simplify the numerical calculation, we use a sharp cutoff at the frequency $\wc$, thus considering a finite number of field modes in the cavity.

Using the dressed ground state (\ref{dressedstata1D}) we can also evaluate the expectation value of the renormalized field correlation function for the scalar field, that in a successive part of this Section we will use to extend our results for the scalar energy density to the case of the one-dimensional electromagnetic field. Writing explicitly space and time components, the renormalized scalar field propagator on the dressed vacuum state (\ref{dressedstata1D}) is given by \cite{ArmataPassante15}
\begin{eqnarray}
G_R(x,t;x',t') &=& \langle g| \phi(x,t)\phi(x',t')| g\rangle_{bc} - \langle\{0_r\}|\phi(x,t)\phi(x',t')|\{0_r\}\rangle_{un}
\nonumber \\
&=& \left(\sum_p \frac{\hbar c^2}{L_0\omega_p} e^{-i\omega_p(t-t')}\sin(k_px)\sin(k_px')-\int\frac{dp}{2\pi} \frac{\hbar c^2}{2\omega_p} e^{-i\omega_p(t-t')}e^{ik_p(x-x')}\right)
\nonumber \\
&+& 8\sum_{j\ell r}
\frac {\hbar c^2}{L_0 (\wj \omega_r )^{1/2}}D_{j\ell}D_{\ell r}
\left[\cos(\omega_jt-\omega_rt')\right] \sin(k_j x)\sin(k_rx') ,
\label{1Dscalarpropagator}
\end{eqnarray}
where subscripts {\it bc} and {\it un} refer respectively to the bounded (Dirichlet) and unbounded case. The quantity in the second line of (\ref{1Dscalarpropagator}) coincides with the renormalized propagator of a fixed-wall cavity, while the term in the third line is the change consequent to quantum motion of the mobile wall around its equilibrium position.

\begin{figure}[t!]
\centering
\subfigure[]{
        \label{Figure1}
        \includegraphics[width=0.48\textwidth]{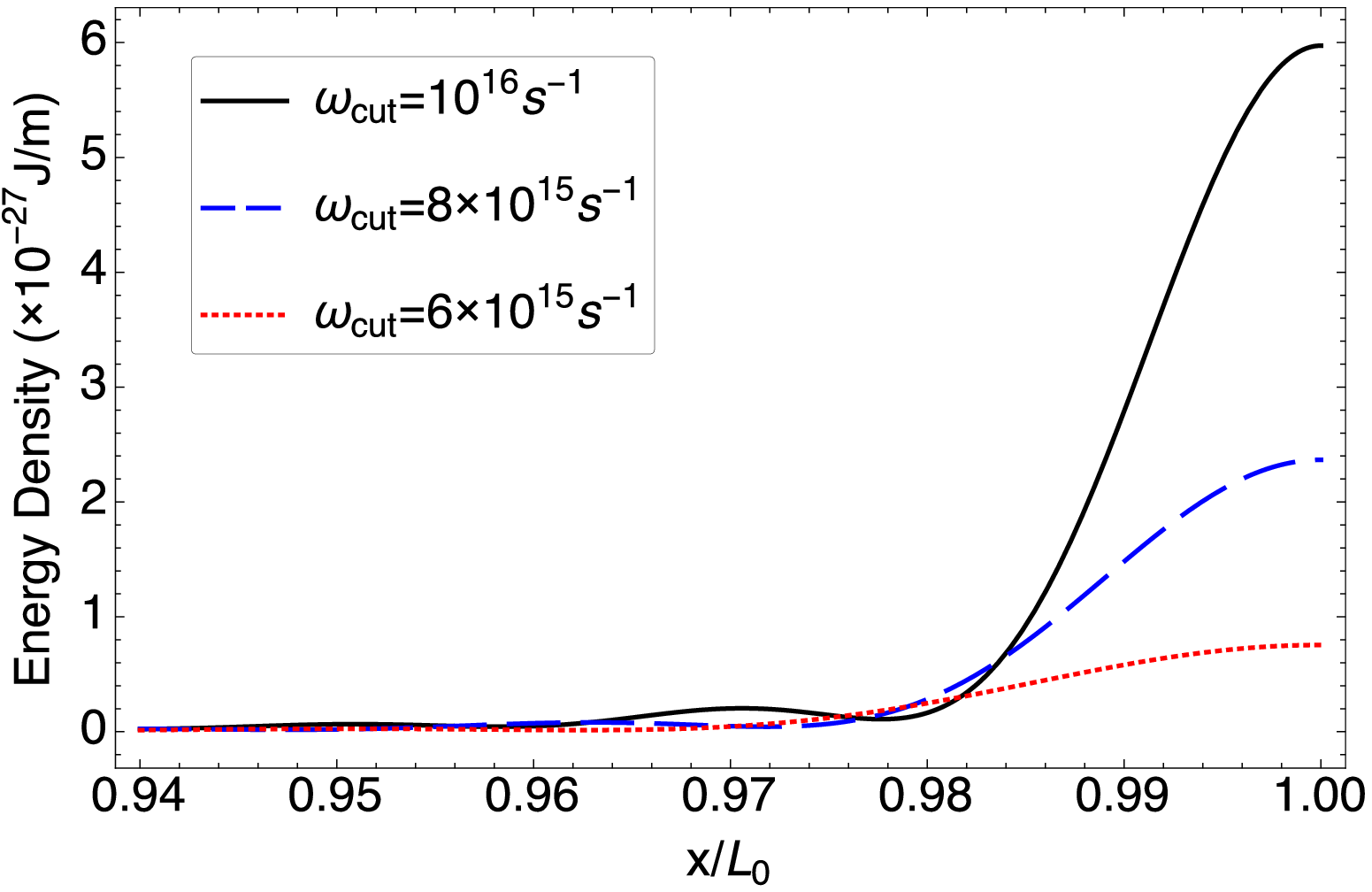}}
\subfigure[]{
        \label{Figure2}
        \includegraphics[width=0.48\textwidth]{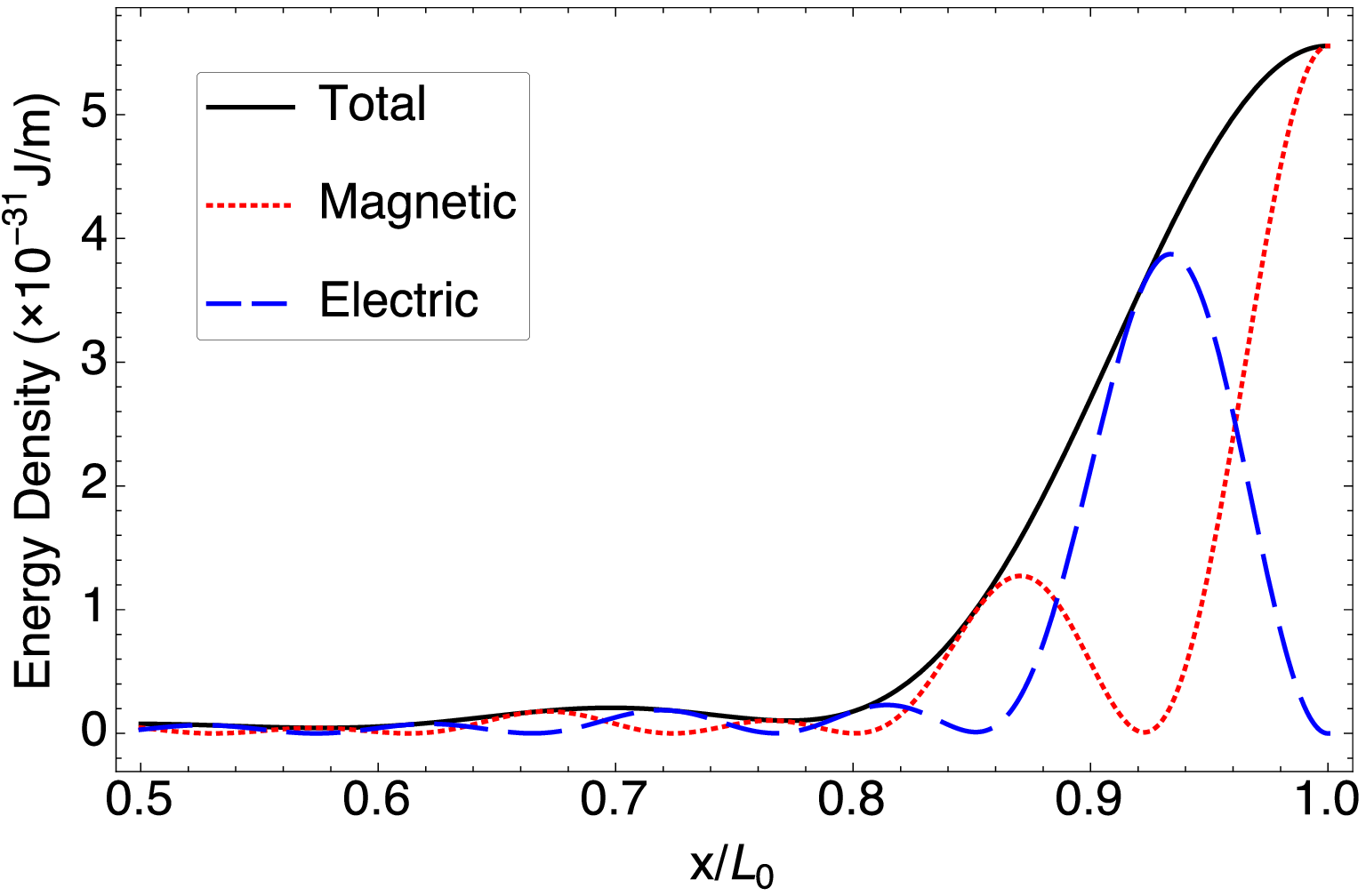}}
\caption{Change of the renormalized field energy density, compared to the static walls case, in the proximity of the moving mirror. In a) the three curves differ for the value of the cutoff frequency:  $\wc = 6 \cdot 10^{15} \, \rm{s}^{-1}$ (red short-dashed line) , $\wc = 8 \cdot 10^{15} \, \rm{s}^{-1}$ (blue long-dashed line) and $\wc = 10^{16} \, \rm{s}^{-1}$ (black continuous line). The field energy density change is concentrated near the mobile wall and grows when the cut-off frequency is increased. b) shows the change of the field energy density (black continuous line), and its electric (blue dashed line) and magnetic (red dotted line) components, due to the motion of the wall. The cutoff frequency has been set to $\wc = 10^{15} \rm{s}^{-1}$. In all plots we have used: $\wo = 10^5 \, \rm{s}^{-1}$, $L_0 = 10 \, \mu \rm{m}$, $M=10^{-11}\, \rm{kg}$.}
\label{Figure}
\end{figure}

Figure \ref{Figure1} shows the change of the renormalized energy density of the field (\ref{averenergydensity1D}) in the proximity of the mobile wall, for three different cut-off frequencies,
specifically $\wc = 10^{16} \, \rm{s}^{-1}, \, \, \wc = 8 \cdot 10^{15} \, \rm{s}^{-1}, \, \, \wc = 6 \cdot 10^{15} \, \rm{s}^{-1}$. This figure clearly shows that the motion of the mobile wall mainly affects the field energy density in its very proximity and is more relevant increasing the cut-off frequency. The values  for the physical parameters used are
$\wo = 10^5 \, \rm{s}^{-1}$, $L_0 = 10 \, \mu \rm{m}$ and $M=10^{-11}\, \rm{kg}$, which are the typical values of commercial microelectromechanical systems (MEMS). It should be noted that the effect scales as $M^{-1}$.  Although in the limit of infinite upper cut-off frequency the energy density diverges at the boundaries \cite{SopovaFord02}, it is possible to shows that averaging over the probability distribution of the position of the mobile wall around its equilibrium position at $L_0$, smears out the divergence \cite{FordSvaiter98,BartoloButera15}.

Using the known relation between the electromagnetic vacuum stress tensor and the renormalized scalar propagator \cite{DeutschCandelas79}, and the possibility to express the electromagnetic field propagator in terms of that for the scalar field \cite{BrownMaclay69}, our results can be generalized to the case of the one-dimensional electromagnetic field. By assuming a one-dimensional model with the cavity length along the $x$ axis, magnetic and electric fields respectively along $y$ and $z$, the renormalized electric and magnetic energy densities result,
\begin{eqnarray}
\langle E_z^2(x) \rangle &=& \frac 1{c^2} \lim_{(x',t') \to (x,t)} \partial_t \partial_{t'}\langle G_R(x,t;x',t') \rangle = \langle E_z^2(x) \rangle_0 + \langle E_z^2(x) \rangle_1 ,
\label{FluctE} \\
\langle B_y^2(x) \rangle &=&  \lim_{(x',t') \to (x,t)} \partial_x \partial_{x'}\langle G_R(x,t;x',t') \rangle = \langle B_y^2(x) \rangle_0 + \langle B_y^2(x) \rangle_1 ,
\label{FluctB}
\end{eqnarray}
where the subscript $0$ indicates the fixed-wall contribution, while $1$ the modification induced by the quantum fluctuactions of the boundary.

Using (\ref{1Dscalarpropagator}), (\ref{FluctE}) and (\ref{FluctB}), we obtain an analytical expression of the average value of the contribution of the wall's motion to the field energy density in the cavity,
$[\langle E_z^2(x) \rangle_0 + \langle E_z^2(x) \rangle_1+\langle B_y^2(x) \rangle_0 + \langle B_y^2(x) \rangle_1]/2$, after introducing an upper cut-off frequency.
Figure \ref{Figure2}  shows the energy density as a function of the position in the proximity of the mobile wall, as well as its electric and magnetic components, using
$\wo = 10^5 \, \rm{s}^{-1}$, $L_0 = 10 \, \mu \rm{m}$, $M=10^{-11}\, \rm{kg}$ and $\wc = 8 \cdot 10^{15} \, \rm{s}^{-1}$. As it can be inferred from the figure, also in the electromagnetic case, the changes are significant mainly close to the mobile wall. This is indeed expected because the pairs of virtual photons emitted by the mobile cavity wall remain confined close to the wall.

The electric and magnetic energy densities can be probed by using a non-dispersive electrically and magnetically polarizable body placed in the cavity at position $x_0$, whose interaction energy with the field fluctuations is given by \cite{PassantePower98}
\begin{equation}
\Delta E_{int}= -\frac 12 \alpha_E \langle E^2(x_0) \rangle -\frac 12 \alpha_M \langle B^2(x_0) \rangle ,
\label{interenergypol}
\end{equation}
where $\alpha_E$ and $\alpha_M$ are respectively the static electric and magnetic polarizability of the probe. Therefore the changes of the electric and magnetic energy densities induced by the wall motion can be measured through an appropriate polarizable body, allowing to detect the effect we have obtained. These results can be also generalized to a massless scalar field in a three-dimensional cavity with a mobile wall \cite{ArmataPassante15}.

\section{Resonance interaction between two entangled atoms in an external environment}
\label{PBG}

We now briefly review some recent results we have obtained on the resonance interaction (mediated by the electromagnetic radiation field) between two identical atoms in the presence of an external environment, specifically a photonic crystal. In the multipolar coupling scheme and in the dipole approximation, our system of two identical atoms A and B interacting with the quantum electromagnetic field is described by the following Hamiltonian \cite{CompagnoPassante95}
\begin{equation}
H = H_A +H_B +\sum_{\bk j} \hbar \wk a_{\bk j}^\dagger a_{\bk j} - \boldsymbol\mu_{A} \cdot \bE ({\bf r}_A) - \boldsymbol\mu_{B}\cdot \bE ({\bf r}_B) ,
\label{Hamiltonian-em}
\end{equation}
where $H_A$ and $H_B$ are the Hamiltonian of atoms A and B with dipole moment operators $\boldsymbol\mu_{A}$ and $\boldsymbol\mu_{B}$ and position ${\bf r}_A$ and ${\bf r}_B$, respectively. We use the two-level approximation with one atom in the excited state $|{e}\rangle$ and the other in the ground $|{g}\rangle$, whose energy difference is $\hbar \wa$: we consider the atoms in their symmetric or antisymmetric state and the field in the vacuum
\begin{equation}
| \psi \rangle_\pm = \frac 1{\sqrt{2}} \left( | g_A, e_B, \{ 0_{\bk j} \} \rangle \pm | e_A, g_B, \{ 0_{\bk j} \rangle \} \right) ,
\label{state}
\end{equation}
The excitation is thus shared between the two atoms. These are also called superradiant (symmetric) and subradiant (antisymmetric) states, and the effect of a perfectly conducting plate on their spontaneous decay rate has been recently considered as a function of the distance of the two atoms from the plate \cite{PalacinoPassante17}. Due to the interaction term in (\ref{Hamiltonian-em}), the state (\ref{state}) is not an eigenstate of the Hamiltonian and we can evaluate the resulting energy shift by second-order perturbation theory, assuming a weak coupling regime. In general the expression of the second-order energy shift depends on the distance between the atoms and gives rise to the so-called resonance interaction energy, thus to a force between the atoms. In the vacuum space, the resonance interaction, due to the exchange of one real or virtual photon between the atoms, is given by \cite{IncardoneFukuta14,Salam10}
\begin{eqnarray}
&\ & \Delta E_\pm^{vac} (r) = \mp (\mu_A^{ge})_i (\mu_B^{eg})_j \left( -\nabla^2 \delta_{ij} + \nabla_i \nabla_j \right) \frac {\cos (k_ar)}{r} \nonumber \\
&=& \pm \frac {(\mu_A^{ge})_i (\mu_B^{eg})_j}{r^3}\left[
\left( \delta_{ij}- 3\hat{{\bf r}}_i \hat{{\bf r}}_j \right) \left( \cos (k_ar) + k_a r \sin (k_ar) \right)
-\left( \delta_{ij}- \hat{{\bf r}}_i \hat{{\bf r}}_j \right) k_a^2 r^2 \cos (k_ar)
\right] ,
\label{interenergyvacuum}
\end{eqnarray}
where $k_a=\wa /c$ and ${\bf r}= {\bf r}_B - {\bf r}_A$ is the distance between the two atoms, and the upper/lower sign refers to the symmetric/antisymmetric state, respectively. In vacuum space, the resonance interaction exhibits a behaviour as $r^{-3}$ in the {\it near} zone $r \ll k_a^{-1}$ and as $r^{-1}$ in the {\it far} zone $r \gg k_a^{-1}$. It is thus a very long range interaction which can be relatively intense, as it is a second-order effect, contrarily to the fourth-order van der Waals and Casimir-Polder dispersion interactions between two atoms. However, in order to observe this interaction, it is necessary to preserve the fragile entangled state (\ref{state}) for a sufficiently long time, and this can be a formidable task from the experimental point of view.

The resonance interaction energy, similarly to other radiative processes, can be changed and tailored through external boundaries or a structured environment. An important example of a structured environment is a photonic crystal, that is a periodic array of dielectric slabs with different refractive index yielding a photonic bandgap, where the photonic density of states vanishes \cite{Yablonovitch87,ElGanainyJohn13,John87,AngelakisKnight04}.
Photonic crystals are a powerful tool to manipulate and tailor many radiative processes, even in the strong-coupling regime \cite{LiuHouch17}.
In our analysis we have considered both the one-dimensional case and the isotropic three-dimensional case, where, in the latter, the one-dimensional dispersion relation is assumed to be valid, regardless of the photon propagation direction. In both cases the photonic density of states outside the forbidden gap, and in the proximity of its edges, can be approximated by a quadratic dispersion relation (effective mass approximation),
$\wk = \omega_g \mp A(k-k_0)^2$, where $\omega_g$ can be the lower gap frequency $\omega_l$ or the upper one $\omega_u$, occurring at the gap wavenumber $k_0$, A is a positive constant and the minus or plus sign applies respectively below the lower edge or above the upper edge of the gap. The quantities $\omega_g = \omega_l \, (\omega_u), k_0, \, A$ depend on the physical properties of the crystal, for example periodicity, dielectric constants, etc. This dispersion relation yields a change of the photon density of states and thus a possible modification of any radiation-mediated physical process: in the one-dimensional and isotropic three-dimensional cases, the density of states vanishes inside the bandgap and has a strong peak in the regions external to the gap and in the proximity of its edges (see \cite{LambropoulosNikolopoulos00}, for example).

We now report some of our results on the resonance interaction between two two-level quantum emitters (atoms or quantum dots) embedded in an isotropic three-dimensional photonic crystal, in two specific cases: (i) when their transition frequency is outside the gap and not far from its upper edge \cite{IncardoneFukuta14}; (ii) when it is inside the forbidden gap and in the proximity of its lower edge \cite{NotararigoPassante16}. We use in both cases the effective mass approximation with the appropriate interpolation $\omega_l = Ak_0^2$, which is necessary to obtain the correct linear dispersion relation for very low-frequency photons, i.e. having wavelength larger than the periodicity of the crystal: in this case, indeed, the photon propagation is essentially the same as in free space. Both results have been obtained by second-order perturbation theory.

In the case (i), we obtain
\begin{equation}
\Delta E_{\pm} = \pm \frac {\omega_u}{2\sqrt{A(\wa - \omega_u)}} (\mu_A^{ge})_i (\mu_B^{eg})_j \left( -\nabla^2 \delta_{ij} + \nabla_i \nabla_j \right) \frac {\cos (k_0 r)}{k_0r} .
\label{interenergyPC1}
\end{equation}

Comparison of (\ref{interenergyPC1}) with (\ref{interenergyvacuum}) shows that the interaction energy scales with the distance as $r^{-1}$ asymptotically ($r \gg k_a^{-1} \sim k_0^{-1}$), as in the vacuum space; however in the photonic crystal case there is the extra numerical factor $\omega_u/(2\sqrt{A(\wa -\omega_u)})$. When the transition frequency is close to the edge of the gap, but not so close to require a nonperturbative calculation, this factor can provide a significant increase of the resonance interactions energy and force, up to a factor $10^3$ for realistic values of the parameters involved \cite{IncardoneFukuta14}. It must be said, however, that in such situations also the spontaneous emission rate is increased, making the correlated state more fragile and difficult to preserve in time. Similar results are obtained when $\wa$ is below the lower edge $\omega_l$ of the gap.

On the other hand, in the case (ii), that is when the atomic transition frequency $\wa$ is inside the gap and in the proximity of its lower edge $\omega_l$, an explicit evaluation of the resonance interaction energy yields a scaling with the distance as $r^{-2}$ in the far zone, rather than the $r^{-1}$ behaviour obtained in the vacuum space. Using typical values for the parameters of a photonic crystal and an optical transition frequency of the two identical atoms, we find that, although the interaction energy can be reduced asymptotically with respect to the free-space case, it is still significant and much stronger than dispersion interactions between atoms  \cite{NotararigoPassante16}. However, contrarily to the case (i), spontaneous emission is strongly suppressed by the presence of the photonic crystal, making more stable the entangled state  (\ref{state}), thus improving the possibility to detect the elusive resonance interatomic interaction.

Another structured environment we have recently investigated to control the resonance interatomic interaction is a cylindric metallic waveguide, when the two entangled atoms are placed on the axis of the guide. In this case the main difference is the presence of a lower cut-off frequency of the guide, directly related to its diameter.  This yields an asymptotical exponential behavior of the interaction with the distance when the atomic transition frequency is below the waveguide's cut-off frequency \cite{FiscelliRizzuto17}.

All the results reported in this section prove that a structured environment, for example a photonic crystal or a waveguide, allows to control and tailor the resonance interaction between two entangled atoms, permitting to increase or reduce the strength of the interaction, as well as controlling the collective spontaneous emission rate.

\section*{Conclusions}
\label{Conclusions}
In this paper we have reviewed some recent results related to the effect of boundaries on several radiative processes, in particular a fluctuating reflecting plate or a photonic crystal. We have first considered a one-dimensional model of a massless scalar field in a cavity with a mobile wall, whose degrees of freedom have been treated quantum mechanically, thus allowing the possibility of quantum fluctuations of its position. We have then extended this model to the case of the one-dimensional electromagnetic field. We have discussed how the position fluctuations of the wall affect the field energy density in the cavity and the related Casimir interaction energies with a polarizable body. We have found significant changes in the very proximity of the mobile wall, which are particularly relevant for masses of the wall as low as those currently realizable in state-of-the-art optomechanics experiments. Successively, we have considered the resonance interaction energy between two atoms or quantum dots embedded in a photonic bandgap material such as a photonic crystal, showing how the presence of the crystal significantly changes the features of the interaction, depending on the position of the atomic transition frequency with respect to the edges of the bandgap. We have also discussed how this setup is promising in the perspective of the direct observation of the elusive quantum resonance interaction between atoms. Finally, we have briefly mentioned preliminary results obtained for the resonance interaction when the two quantum emitters are placed within a cylindrical metallic waveguide.

\ack
F.A. acknowledges the Marie Curie Actions of the EUÕs 7th Framework Program under REA [grant number 317232] for their financial support. S.B. acknowledges support from the EPSRC CM-CDT Grant No. EP/403673X/1.

\end{document}